\documentclass[longbibliography,pre,preprint,singlecolumn,showpacs]{revtex4-1}
\usepackage{amsmath}
\usepackage{graphicx}
\usepackage{bm}

\begin{document}
\title{Dynamical Effective Field Model for Interacting Ferrofluids: I. Derivations for homogeneous, inhomogeneous, and polydisperse cases}

\author{Angbo Fang}
\affiliation{School of Physics and Electronics, North China University of Water Resources and Electric Power, Zhengzhou 450011, China}
\date{\today}

\begin{abstract}
Quite recently I have proposed a nonperturbative dynamical effective field model (DEFM) to quantitatively describe the dynamics of interacting ferrofluids.  Its predictions compare very well with the results from Brownian dynamics simulations.   In this paper  I put the DEFM on firm theoretical ground by deriving it within the framework of dynamical density functional theory (DDFT), taking into account nonadiabatic effects.  The DEFM is generalized to inhomogeneous finite-size samples for which the macroscopic and mesoscopic scale separation is nontrivial due to the presence of long-range dipole-dipole interactions.  The demagnetizing field naturally emerges from microscopic considerations and is consistently accounted for.  The resulting particle dynamics on the mesoscopic scale only involves macroscopically local quantities such as local magnetization and Maxwell field.  Nevertheless, the local demagnetizing field essentially couples to magnetization at distant macroscopic locations.  Thus, a two-scale parallel algorithm, involving information transfer between different macroscopic locations, can be applied to fully resolve particle rotational dynamics in an inhomogeneous sample.   I also derive the DEFM for polydisperse ferrofluids, in which the dynamics of particles belonging to different species can be strongly coupled to each other.  I discuss the underlying assumptions in obtaining a thermodynamically consistent polydisperse magnetization relaxation equation, which is of the same generic form as that for monodisperse ferrofluids. The theoretical advances presented in this paper are important for both qualitative understanding and quantitative modeling of ferrofluid dynamics.
\end{abstract}
\maketitle
\section{Introduction}
Ferrofluid~\cite{Rosen:1985} is a colloidal suspension formed by dispersing single-domain ferromagnetic nanoparticles into a nonmagnetic liquid carrier.  From the academic point of view, it is a prototype for many dipolar fluids (either molecular or colloidal) with long-range dipole-dipole interactions (DDI).  From the application point of view, it is a highly functional soft material with controllable and tunable physical (mechanical, thermal, magnetic, optical, etc.) properties.  In addition to well-developed applications of ferrofluids in industry and biomedicine,  new applications keep emerging~\cite{torres2014recent}.   Either for our fundamental understanding of this prominent material or for optimizing its performance in a wide range of applications, it is a key issue to quantitatively describe the dynamic magnetic response of ferrofluids. On the macroscopic time and length scales relevant to most situations,  this is achieved via the magnetization relaxation equation (MRE).

Historically, the first MRE (in the context of molecular liquids carrying electric dipoles) is due to Debye~\cite{Debye:1929}.  He started from the Smoluchowski equation (SE) describing Brownian rotational motion of a spherical rigid dipolar particle under a constant field.  In ferrofluid context, within the rigid-dipole approximation a particle carries a magnetic moment $\bm{\mu}$ whose direction is locked to its orientational director $\bm{e}$, i. e.,  $\bm{\mu}=\mu \bm{e}$.
If we denote $W(\bm{e}, t)$ the orientational distribution function (ODF), then the single-particle SE reads
\begin{equation}
2\tau^0_r \frac{dW}{dt}= \frac{1}{k_B T} \widehat{\mathcal{R}}_e W \widehat{\mathcal{R}}_e \left[k_B T \ln W - \mu_0 \mu \bm{e}\cdot \bm{H} \right],
\end{equation} 
where $\bm{H}$ is the external magnetic field, $\mu_0$ is the vacuum magnetic permeability, $k_B$ is the Boltmann constant, $T$ is the absolute temperature, and $\widehat{\mathcal{R}}_e =\bm{e}\times \partial/{\partial\bm{e}}$  is the infinitesimal rotation operator or gradient operator on the surface of unit sphere.  For brevity $\mu_0$ will be absorbed into the thermal energy and not explicitly shown in later discussions.  In Eq.~(1),  $\tau^0_r$ is  the Brownian (or Debye's) rotational relaxation time of a tagged particle in a infinitely dilute suspension.  According to the Stokes-Einstein-Debye relation, it is given by
\begin{equation}
\tau^0_r = \frac{3\eta_s V_p}{k_B T},
\end{equation} 
where $\eta_s$ the shear viscosity of the solvent and $V_p$ the hydrodynamic volume of the tagged particle.
Debye considered the case when the external field is small so that Eq.~(1) can be linearized around the unpolarized equilibrium state.  This leads to the well-known Debye relaxation equation.  In 1974 Martsenyuk, Raikher and Shliomis (MRSh)~\cite{MRSh:1974} developed a new nonlinear MRE by manipulating SE (1) with an effective field ansatz.  The MRSh model applies to regimes far away from equilibrium.  Nevertheless, it is still restricted to ideal ferrofluids, i. e., an ensemble of noninteracting monodisperse paramagnetic particles.

However, in typical ferrofluids particles  interact with each other via steric repulsions and long-range magnetic DDIs.
They are not negligible but usually of crucial importance in determining static and dynamic properties.   Hence in general we have to start from an $N$-particle SE
to take care of inter-particle correlations.   To make the problem trackable, it is desirable to reduce it to an effective single-particle SE, with the effects of particle interactions approximately described by a one-body mean-interaction potential.

There are two important issues related to this reduction, encountered in solving ubiquitous many-body problems.  The first, of general theoretical interest, is to obtain a single-particle SE in closed form, under some physically transparent assumptions and without losing generality.
A well-accepted formulism called the classical dynamical density functional theory (DDFT),  has been proposed by Marconi and Tarazona~\cite{Marconi1999ddft, Tarazona2000ddft} and further reformulated and revised by others~\cite{Archer2004ddft, Lowen2007ddft, Goddard2012ddft, Schmidt2013power}, providing a general route to this many-to-one reduction for a collection of interacting Brownian particles. For a comprehensive review of DDFT, see Ref.~\citenum{Lowen2020review}.  The second issue, of practical significance, is to find an analytical and quantitatively reliable approximation for the mean-interaction potential that sufficiently accounts for
the effects of inter-particle correlations.  This has to be system-specific and often relies on our insights on the underlying physics.

Quite recently, I have proposed a nonperturbative dynamical effective field model (DEFM)~\cite{FangSM2020} for interacting ferrofluids, in which the mean-interaction potential is assumed to be produced by a non-equilibrium excess effective field that is self-consistently determined as  a function of the instantaneous magnetization.  This crucial step is motivated by observations on previous perturbative models.   By comparing the theoretical predictions and Brownian dynamics (BD) simulations on the dynamic magnetic susceptibilities (DMS) under both zero and finite bias fields,  the DEFM is demonstrated to be of quantitative reliability for concentrated, strongly interacting and polydisperse ferrofluids.

Nevertheless, to eliminate the heuristic nature of the model and put it on firm ground, it is highly desirable to provide a decent derivation.  This will not only give us more confidence in applying it to various problems of practical importance,  but also clarify the underlying assumptions and delineate its validity regimes.  Furthermore,  the original model is formulated for unbounded and homogeneous ferrofluids, but real samples are of finite size and inhomogeneous.  Thus, it has to be revised to apply to the inhomogeneous case.
In addition, it is nontrivial to establish the DEFM for polydisperse interacting ferrofluids, in which we have an ODF for each species of particles and they are coupled together.
I will tackle these important issues in Paper I.

In DDFT, to obtain a closed single-particle SE,  a one-step adiabatic approximation is made on the pair correlation functions.  Such an approximation implies a time scale coarse-graining, via which the dynamic evolution of inter-particle correlations is no longer explicitly accounted for.  As taught by statistical mechanics, effects of small-scale fluctuations are encoded in the corresponding transport coefficients on a large scale.   Therefore, it is expected that the neglected dynamic correlations, should manifest themselves by renormalizing the relevant transport coefficients in the effective single-particle SE.
Traditional DDFT schemes, however, have either completely or partially discarded the effects of dynamic correlations, even though it is  essential for most concentrated and strongly interacting systems. It is understood that an effective single-particle dynamical model describes a dressed other than bare particle .  The characteristic relaxation time of the former can be strikingly different from the latter, due to the integrated effects of short-time dynamic correlations.

I devote Paper II to bridging macroscopic and microscopic relaxation times in interacting ferrofluids, thereby identifying the proper characteristic relaxation time in the DEFM.  Simple definitions for correlation factors are proposed to describe the effects of static and dynamic inter-particle correlations on macroscopic relaxation dynamics, respectively.  Via both factors DMS is elegantly and analytically expressed in terms of (Debye's) frequency-dependent single-particle orientational susceptibility.  I show how to determine the dynamic correlation factor from DMS measurements.  For typical monodisperse ferrofluids, an empirical formula is proposed for it. My theoretical predictions compare well with results from BD simulations.

Paper I is structured as follows.  In Sec. II a derivation of the DEFM is provided for homogeneous monodisperse ferrofluids, within the framework of DDFT for classical fluids.
The underlying assumptions are carefully discussed.  Sec. III develops the revised DEFM applicable to inhomogeneous ferrofluid samples. The demagnetizing field naturally emerges from the uncorrelated part of inter-particle DDIs.  The effects of inter-particle correlations are consistently accounted for via a correlation-induced effective field.  In Sec. IV the DEFM is established for polydisperse interacting ferrofluids, from which the polydisperse MRE is derived.  Conclusions are drawn in the last section.

\section{DEFM for homogeneous monodisperse Ferrofluids}
\subsection{DDFT for rotational dynamics}
I will derive the DEFM model via two steps.  In this subsection the $N$-particle SE is reduced to a closed single-particle DDFT equation.  The effects due to short-time dynamics, induced by HIs or fluctuating interactions among particles,  are absorbed into the effective diffusion coefficient.  The ensemble of bare particles are mapped to an ensemble of dressed particles without short-time dynamic correlations, with equivalent dynamics on the slow time scale.   Then, in the second subsection,  a dynamical effective field approximation is further employed to render the DDFT equation fully explicit, leading to the DEFM model.

I start by considering $N$ identical spherical particles dispersed in a viscous solvent, occupying total volume $V$, maintained at temperature $T$, and subject to an external potential.  On the time and length scales relevant to us,  the solvent can be treated as a structureless and continuous medium. On the diffusive regime ($t \gg \tau_p$, with $\tau_p$ the characteristic time beyond which linear and angular momenta of colloidal particles are overdamped), we can employ the $N$-particle probability density $P(\bm{r}^N, \bm{e}^N, t)$ to describe the distribution of particles with different positions
$\bm{r}^N=(\bm{r}_1, ..., \bm{r}_N)$ and orientations $\bm{e}^N=(\bm{e}_1, ..., \bm{e}_N)$.   The time evolution of  $P(\bm{r}^N, \bm{e}^N, t)$ is governed by the $N$-particle SE
\begin{equation}
\frac{\partial P(\bm{r}^N, \bm{e}^N, t)}{\partial t} = \widehat{\mathcal{L}}  P(\bm{r}^N, \bm{e}^N, t),
\end{equation}  
with the Smoluchowski operator defined by
\begin{equation}
\widehat{\mathcal{L}} = \sum^N_{i=1}
\left\{ D_0
 \nabla_{\bm{r}_i} \cdot \left[\nabla_{\bm{r}_i} + \beta \nabla_{\bm{r}_i} U(\bm{r}^N, \bm{e}^N, t) \right]
+ D^0_r \widehat{\mathcal R}_i \cdot \left[\widehat{\mathcal R}_i + \beta \widehat{\mathcal R}_i U(\bm{r}^N, \bm{e}^N, t) \right]
\right\},
\end{equation} 
in which $\beta=1/k_B T$ and $D_0$ and $D^0_r$ are respectively the translational and rotational single-particle diffusion coefficients.
Including the effect of hydrodynamic interactions (HI) would account for cross diffusions between different particles. For brevity of formulation I will not consider HIs for the moment. Furthermore, $\nabla_{\bm{r}_i}$ is the gradient operator with respect to $\bm{r}_i$ and  $\widehat{\mathcal R}_i= \bm{e}_i \times \nabla_{\bm{e}_i}$ is the rotation operator acting on the orientation vector $\bm{e}_i$ located on a unit sphere surface.  Usually, the total potential energy $U(\bm{r}^N, \bm{e}^N, t)$
are of the following form:
\begin{equation}
U(\bm{r}^N, \bm{e}^N, t) = \sum_{i=1}^N v_{ext}(\bm{r}_i, \bm{e}_i, t) + \frac{1}{2} \sum_{j\ne i}^N \sum_{i=1}^N v_2(\bm{r}_i, \bm{r}_j, \bm{e}_i, \bm{e}_j),
\end{equation} 
where $v_{ext}$ is the one-body external potential and $v_2$ is the pair potential.

For the problem to be trackable, we may reduce the dimension by considering the $n$-body ($n < N$) density distribution function defined by
\begin{equation}
P^{(n)}(\bm{r}^n, \bm{e}^n, t) = \frac{N!}{(N-n)!}\int d\bm{r}_{n+1}\oint d\bm{e}_{n+1} ...\int d\bm{r}_{N}\oint d\bm{e}_{N}
P(\bm{r}^N, \bm{e}^N, t).
\end{equation}  
To obtain the simplest description of particle dynamics at the mesoscopic level,  we can integrate $P(\bm{r}^N, \bm{e}^N, t)$ with $N\int d\bm{r}_2\oint d\bm{e}_2...\int d\bm{r}_N\oint d\bm{e}_N$ to obtain
the single-particle density $P^{(1)}(\bm{r}, \bm{e}, t)$.
The $N$-particle SE (3) can be integrated to yield the equation of motion for $P^{(1)}$. However, this equation is not closed since it still couples with $P^{(2)}$, whose equation of motion in turn depends on $P^{(3)}$.   And so on.  This is known as the Bogoliubov-Born-Green-Kirkwood-Yvon (BBGKY) hierarchy.

Now I focus on monodisperse ferrofluids and assume the sample is uniform in positional space.
The external potential is given in the form of $v_{ext}(\bm{e}, t)$, i. e., it only couples to particle orientations. The interaction potential between a pair of particles  is $v_2(\bm{r}-\bm{r}', \bm{e}, \bm{e}') = v_s(|\bm{r}-\bm{r}'|) + v^{dd}(\bm{r}-\bm{r}', \bm{e}, \bm{e}')$, with $v_s$ describing the isotropic short-range repulsive interaction and $v^{dd}$ describing the long-range DDI, respectively.  The translational degrees of freedom for colloidal particles relax in a much shorter time than the rotational degrees of freedom.  Hence in the rotational diffusion regime when particle positions are overdamped, we have $P^{(1)}(\bm{r}, \bm{e}, t) = \rho W(\bm{e}, t)$, with $\rho= N/V$ the particle number per unit volume and $W(\bm{e}, t)$ the single-particle ODF.
Furthermore,  we have $P^{(2)}(\bm{r}_1, \bm{r}_2, \bm{e}_1, \bm{e}_2, t) = \rho^2 W(\bm{e}_1, t)W(\bm{e}_2, t)
g^{(2)}(\bm{r}_1- \bm{r}_2, \bm{e}_1, \bm{e}_2, t)$, with $g^{(2)}$ the pair correlation function (PCF).

The reduced single-particle rotational SE reads:
\begin{equation}
\frac{\partial W(\bm{e}, t)}{\partial t} = \frac{D^0_r}{k_B T } \widehat{\mathcal R} W(\bm{e}, t) \cdot \widehat{\mathcal R}  \left[k_B T \ln{W(\bm{e}, t)} +  v_{ext}(\bm{e}, t) + \phi_{int}(\bm{e}, t) \right],
\end{equation}  
where $\phi_{int}$ is the non-equilibrium excess chemical potential due to interactions of a representative particle with all other particles.   It is given by
\begin{equation}
\phi_{int}(\bm{e}, t)= \int d\bm{r} \oint d\bm{e}' g^{(2)}(\bm{r}, \bm{e}, \bm{e}', t)W(\bm{e}', t) v_2(\bm{r}, \bm{e}, \bm{e}').
\end{equation}  
It is noted that the first and second term in the bracket of Eq.~(7) can be identified with the thermal chemical potential due to the ideal-gas entropic contribution, denoted by $\phi_{id}[W(\bm{e}, t)] = k_B T \ln{W(\bm{e}, t)}$, and the external chemical potential, denoted by $\phi_{ext}\equiv v_{ext}$, respectively.

Because $\phi_{int}$ depends on $g^{(2)}$, Eq.~(7) is not closed.
Marconi and Tarazona~\cite{Marconi1999ddft, Tarazona2000ddft} have proposed a closure scheme called ``dynamical density functional theory" (DDFT), based on the adiabatic approximation of inter-particle correlations. Denoting $\tau_g$ the characteristic decay time of $g^{(2)}$, it essentially amounts to
perform a time-averaging of Eq.~(7) over a duration $\Delta t \gg \tau_g$.  For many systems subject to a slowly-varying external potential,  single-particle distribution varies much slowly than pair and higher-order correlations.  This time scale separation greatly simplifies the time-averaging procedure.
At any fixed time $t$ on the coarse-grained time scale, we may simply replace $g^{(2)}(\bm{r}, \bm{e}_1, \bm{e}_2, t)$
with the static pair correlation function $g^{eq}_t(\bm{r}, \bm{e}, \bm{e}')$ for
a suitable equilibrium reference system whose ODF is given by $W^{eq}_t(\bm{e})\equiv W(\bm{e}, t)$.
Here and below, the subscript ``t" denotes an adiabatic parameter labeling the ``instantaneous" (on the coarse-grained time scale) equilibrium reference system.

Such a reference system can always be prepared by a modified external potential $\widetilde{v}_t(\bm{e})$.
Since it is in equilibrium, the total chemical potential has to be a constant in the orientational space~\cite{Evans:1979}:
\begin{equation}
\phi_{id}[W^{eq}_t(\bm{e})] + \widetilde{v}_t(\bm{e}) + \frac{\delta \mathcal{F}_{int}[W^{eq}_t(\bm{e})]}{\delta W^{eq}_t(\bm{e})}= Const,
\end{equation}  
where $\mathcal{F}_{int}$ is the excess free energy functional  due to particle interactions.

Furthermore, according to the well-known Yvon-Born-Green (YBG) relation~\cite{Hansen2006theory, Gubbins1980structure}, we also have
\begin{equation}
\widehat{\mathcal R} \left[\phi_{id}[W^{eq}_t(\bm{e})] + \widetilde{v}_t(\bm{e})+ \phi^{eq}_t(\bm{e})\right] =0,
\end{equation}  
where $\phi^{eq}_t$ is obtained from Eq.~(8) by replacing $g^{(2)}(\bm{r}, \bm{e}_1, \bm{e}_2, t)$ and $W(\bm{e}, t)$ with $g^{eq}_t(\bm{r}, \bm{e}, \bm{e}')$ and $W^{eq}_t(\bm{e})$, respectively.  Eq.~(10) describes the condition of generalized torque balance on a representative particle.

Combining Eqs.~(9) and (10), it is clear that, if the adiabatic approximation is taken,  then up to an irrelevant constant, the non-equilibrium  excess chemical potential can be expressed  as follows:
\begin{equation}
\phi_{int}(\bm{e}, t) =  \frac{\delta \mathcal{F}_{int}[W(\bm{e},t)]}{\delta W(\bm{e}, t)},
\end{equation}  
which is now decoupled from $g^{(2)}$.  Therefore Eq.~(7) is transformed to a closed effective single-particle SE:
\begin{equation}
\frac{\partial W(\bm{e}, t)}{\partial t} = \frac{D_r}{k_B T} \, \widehat{\mathcal R}_e \, W(\bm{e}, t) \cdot \widehat{\mathcal R}_e  \left[k_B T \ln W(\bm{e}, t) +  
v_{ext}(\bm{e}, t) + \frac{\delta \mathcal{F}_{int}\left[W(\bm{e}, t)\right]}{\delta W(\bm{e}, t)} \right].
\end{equation}  

Now, consider the effects of HIs.  Starting from an $N$-particle SE including HIs, integrating out $N-1$ particles,  and taking the adiabatic approximation, we will obtain a closed equation of motion for the single-particle distribution function.  Such a procedure was first carried out by Rex and L\"{o}wen~\cite{RexLowen2009HI}  and later revised by others~\cite{Donev2014HI},   with respect to translational motion of colloids in dilute suspensions.  To transfer their treatments to rotational dynamics of ferrofluids, Eq.~(7) is rewritten as $\partial W/{\partial t} = - ({D^0_r}/{k_B T}) \widehat{\mathcal R}_e\cdot \bm{J}_0$, where $\bm{J}_0 = - \widehat{\mathcal R}_e \left[\phi_{id}+v_{ext}+\phi_{int} \right]$ the single-particle orientational flux.  For a tagged particle located at $\bm{r}$, to the pair level the far-field HIs will induce two additional fluxes amending $\bm{J}_0$.  One describes the contribution of flow vorticity induced at $\bm{r}$ reflected by another particles at $\bm{r}' \ne \bm{r}$,  while the other arises from the torque acting on another particle at $\bm{r}'$, where a vortice is generated and transmitted to $\bm{r}$.   In dilute suspensions the second flux dominates, which mathematically can be represented as an integral over $\bm{r}'$ (and particle orientation $\bm{e}'$),  of the nonlocal Rotne-Prager tensor multiplied by the torque at $\bm{r}'$ (given by $\widehat{\mathcal R}_{e'}\left[v_{ext}(\bm{e}', t)+ \phi_{int}(\bm{e}', t)\right]$)  and weighted by the probability of finding a pair of particles with configurations $(\bm{r}, \bm{e})$ and $(\bm{r}', \bm{e}')$.  As the latter involves PCF,  another closure condition is needed in addition to the adiabatic expression for $\phi_{int}$. While a consistent adiabatic expression for the time-dependent PCF can in principle determined from the Ornstein-Zernike equation~\cite{RexLowen2009HI},  a desired analytic expression is generally not available.  Moreover, in concentrated suspensions, higher-order HIs have to be included, rendering this explicit scheme even less practical.

If we examine more carefully on how Eq.~(12) is derived from Eq.~(7), it is not harder to find what has been discarded in regular derivations of DDFT equations,
whether HIs are included~\cite{RexLowen2009HI, Donev2014HI} or not~\cite{Marconi1999ddft, Tarazona2000ddft, Archer2004ddft}.
Essentially,  the adiabatic approximation on inter-particle correlations is the consequence of time scale coarse-graining or ensemble averaging.   To perform a dynamic coarse-graining, we should first pick out the appropriate set of slow observables,  which, in Eq.~(7), is just the single-particle ODF.   Then, each term in Eq.~(7) can be coherently coarse-grained by following the projection operator approach~\cite{Espanol2009POM}.  Every involved observable is decomposed as the sum of a fast and a slow component.   Here, supposing $v_{ext}$ is slowly varying, the PCF is the only relevant observable having a fast component.  Denoting $g^{(2)}_0$ and $\widetilde{g}^{(2)}$ respectively as its slow and fast components, the adiabatic approximation is equivalent to discarding $\widetilde{g}^{(2)}$.   What is the consequence of this approximation?

Via the projection operator,  the overall interaction on a tagged particle due to other particles can be divided into two parts: one is a time-averaged contribution well described by quasi-static PCF under the adiabatic approximation, whereas the other is a fluctuating contribution depending on the dynamic evolution of PCF at short times.   It is this fluctuating part that gives rise to additional friction or caging effect distinguishing long-time from short-time diffusivity.  This implies, even if HIs are not included,  for consistency a renormalization of the diffusion coefficient should be carried out,  to describe the time-integrated memory effects due to dynamic correlations at short times.
This point seems not widely captured, probably because most systems studied with DDFT are either dilute or with weak interactions and studies have been focused on the effect of HIs.  The latter are usually treated to pair level and weighted by the adiabatic PCF~\cite{RexLowen2009HI, Donev2014HI}.

While the projection operator approach enables formally tracking the dynamic effects of fast components on renormalization of transport coefficients,  usually it can not yield useful analytic expressions.   This is true even if HIs are excluded.   Therefore, I propose to simplify the formulism by adopting a phenomenological approach,  based on some plausible assumptions.  Firs of all, I assume on an appropriate slow time scale (still much shorter than the hydrodynamic scale but long enough for the adiabatic approximation to hold),   cross diffusions (induced by either hydrodynamic or direct interactions) are negligible. This indicates insignificance of  the integrated memory effects due to  angular velocity correlation between different particles.   Then we can obtain a closed effective single-particle dynamic equation, similar to what Marconi and Tarazona originally achieved. However, unlike their work in which only quasi-static correlations are accounted for, here I keep the integrated effect due to spatially local correlations at short times,
whether it arises from hydrodynamic or direct interactions.   Its sole effect is to renormalize the bare (rotational) self-diffusion coefficient.   This assumption seems to be justified in monodisperse ferrofluids with negligible particle aggregations (see discussions in Paper II).    Second, the total flux inducing the evolution of single-particle ODF is still assumed given by $J_0$ under the adiabatic approximation.  This means  reflective contributions of HIs only give rise to an additional renormalization factor but not an new flux.  Finally, I assume the renormalized rotational self-diffusion coefficient, denoted by $D^*_r$, is a function of two key material parameters of monodisperse ferrofluids.  One is the hydrodynamic volume fraction $\phi\equiv (\pi/6) \rho d^3$, with $d$ the particle diameter; the other is the characteristic strength of DDIs, defined by $\lambda=\mu^2/(4\pi d^3 k_B T)$, with $\mu$ the magnetic moment per particle.  Notably, for hard-sphere colloidal suspensions, this kind of phenomenological treatment~\cite{Hayakawa1995statistical, Royall2007HI} has often been utilized in accounting for HIs in an averaged manner, with the bare diffusion coefficient replaced by the one describing equilibrium long-time dynamics.  The latter is usually a function of $\phi$, whose form, to leading order of $\phi$, can be calculated with perturbation methods~\cite{Jones1989rotational, Cichocki1999lubrication}.  In ferrofluids, whereas HIs give rise to $D^*_r$ a $\phi$ dependence, DDIs can cause $D^*_r$ to depend on both $\lambda$ and $\phi$.   In general, particle packing influences both hydrodynamic and direct interactions and their impacts on particle self-diffusivity are coupled together.

Nevertheless, Medina-Noyola~\cite{Medina1988long} has provided an important insight on long-time translational self-diffusion in concentrated colloidal dispersions.  He argued that the effects of HIs and direct interactions are approximately decoupled, with the former modifying short-time diffusivity but the latter dominating long-time diffusivity due to additional friction arising from  inter-particle collisions.  If this is true for rotational dynamics of ferrofluids, then we may write $D^*_r/ D^0_r = \psi_h(\phi) \psi_c(\phi, \lambda)$, with $\psi_h$ and  $\psi_c$ denoting the effects of HIs and DDIs, respectively.  Then the two factors can be evaluated separately and approximations can be made respectively on short and long time scales,  easing the involved calculations a lot.

Based on the above discussions, phenomenologically, the effective single-particle dynamic equation still assumes the form of Eq.~(12), but with $D^0_r$ replaced by $D^*_r$,  taking care of  the integrated effect of short-time dynamic correlations~\cite{NotePower20201109}.  Conceptually, we may say, by time scale coarse-graining, the original ensemble of bare brownian particles are mapped to an ensemble of  dressed (or quasi-) particles.  Each dressed particle looks like the same as a bare particle except for their distinct self-diffusivity.  Dressed particles absorb all short-time fluctuations and diffuse independent of each other.  But they still interact with each other, couple to external fields, and form quasi-static mesoscopic structure, in exactly the same manner as the bare particles do on a slow time scale.

However, with this phenomenological mapping, we lose the quantitative connection between $D^*_r$ for a dressed particle and $D^0_r$ for a bare particle. This is because the mapping is not explicitly defined,  in contrast to the coarse-graining procedure carried out via the projection operator approach.    The lost connection will be restored in Paper II.

\subsection{Dynamical Effective Field Approximation}

While the bare-to-dressed ensemble mapping along with the DDFT scheme provides a closure for single-particle reorientation dynamics, we still need to construct an accurate expression for $\mathcal{F}_{int}$.  This is especially difficult for strongly interacting systems for which perturbative models are often inadequate.
Nevertheless, for ferrofluids with inter-particle orientational correlations dominated by long-range DDIs, a mean field approximation to $\mathcal{F}_{int}$  seems to provide quite good descriptions of both static and dynamic properties.  The modified  mean field  theories~\cite{Ivanov2001magnetic, Ivanov2007magnetic, Ivanov2017modified}, though of perturbative nature, are highly successful in describing  the equilibrium properties of interacting ferrofluids, either monodisperse or polydisperse.  Quite recently, I have proposed a non-perturbative DEFM~\cite{FangSM2020} for ferrofluids, with the excess chemical potential explicitly given as  a function of the instantaneous magnetization.
It is proven quantitatively accurate for describing the DMS of both monodisperse and bidisperse ferrofluids.  In addition,  quantitative accuracies have been demonstrated for monodisperse ferrofluids driven far away from equilibrium by a finite bias field.   Although  the DEFM originally arises from observations on low-order perturbative models~\cite{MRSh:1974, Zubarev:1998, Fang:2019pof},  I will show below that it can be derived from Eq.~(12) (with $D^0_r$ replaced by $D^*_r$)  with some reasonable assumptions.

Let us consider $\mathcal{F}_{int}(W_t(\bm{e}))$ for the equilibrium reference ferrofluid described by $W_t(\bm{e})$.  Alternatively, we can regard $\mathcal{F}_{int}$ as a function of all-order moments of $W_t(\bm{e})$.  On the coarse-grained macroscopic level,   the magnetization $\bm{M}_t = \rho \mu \oint  d\bm{e}\,  W_t(\bm{e})\, \bm{e}$ is the only relevant thermodynamic variable for typical ferrofluids,  in which  there is no appreciable particle aggregation to form inhomogeneous microstructures such as chains or rings.  Hence we may write $\mathcal{F}_{int}$ as the sum of two decoupled parts, with $\mathcal{F}_0(\bm{M}_t)$
the relevant part depending only on $\bm{M}_t$ and $\mathcal{F}_{\Delta}$ the irrelevant part depending on the fluctuations of all other higher-order moments.   It is further assumed  $\mathcal{F}_{\Delta}$ resulting in zero mean torque on the time scale $\Delta t >> \tau_g$.  Thus,  under the adiabatic approximation, we have, up to an irrelevant  constant,
\begin{equation}
\phi_{int}(\bm{e}, t) = \frac{\delta \mathcal{F}_0\left[ W(\bm{e}, t)\right]}{\delta W(\bm{e}, t)} = -\mu \bm{e}\cdot \bm{H}^{exc}(t).
\end{equation} 
In Eq.~(13) $\bm{H}^{exc}$ is the non-equilibrium excess effective field defined via
\begin{equation}
\bm{H}^{exc}(t) \equiv  - \rho \frac{d \mathcal{F}_0\left(\bm{M}(t)\right)}{d\bm{M}(t)} = \widetilde{H}^{exc} (M_t)  \widehat{\bm{m}}_t ,
\end{equation}  
where $\widehat{\bm{m}}_t\equiv \bm{M}(t) /M(t)$ is the director along the instantaneous magnetization and  $\widetilde{H}^{exc}$ is a scalar function of $M_t$, to be determined below.

Now, substituting Eqs.~(13) and ~(14) into Eq.~(12)  we obtain
\begin{equation}
\frac{\partial W(\bm{e}, t)}{\partial t} = \frac{D^{*}_r}{k_B T} \widehat{\mathcal R} \cdot W(\bm{e}, t) \widehat{\mathcal R} \left[
 k_B T  ln W(\bm{e}, t) + v_{ext}(\bm{e}, t) - \mu \bm{e}\cdot\bm{H}^{exc}(t)  \right].
\end{equation}  

To determine the precise form of $H^{exc}$ as a function of $M(t)$,  I consider the stationary profile of Eq.~(15) when the external potential is resulted from a static magnetic field $\bm{H}_s$.  At this forced equilibrium, the orientation profile is denoted by $W^{eq}(\bm{e})$, the magnetization by $\bm{M}_{eq}=\rho \mu \oint d\bm{e} \, \bm{e} \,W^{eq}(\bm{e})$, and the associated excess effective field by $\bm{H}^{exc}_{eq} \equiv \widetilde{H}^{exc} (M_{eq}) \bm{M}_{eq}/M_{eq}$.
According to Eq.~(10) the generalized torque balance condition reads:
\begin{equation}
k_B T \bm{e}\times \frac{d\ln W^{eq}(\bm{e})}{d\bm{e}} - \bm{e} \times \bm{H}_s - \bm{e} \times \bm{H}^{exc}_{eq} = 0,
\end{equation}  
which can be integrated, yielding
\begin{equation}
W^{eq}(\bm{e}) = Z^{-1} \exp\left[\bm{e} \cdot (\bm{H}_s + \bm{H}^{exc}_{eq})/k_B T\right],
\end{equation}  
with $Z$ a normalization factor.    Notably, $W^{eq}(\bm{e})$  is of uniaxial nature, as learned from studies of ferrofluids
prepared in equilibrium by a static magnetic field~\cite{Ivanov2001magnetic}.  However, in general $W(\bm{e},t)$ (or equivalently, $W^{eq}_t(\bm{e})$) does not necessarily possess this uniaxial nature.  This implies, the external potential $\widetilde{v}_t(\bm{e})$ prescribed to prepare $W^{eq}_t(\bm{e})$,  in general, can not be solely produced by a static magnetic field.  Nevertheless, typical ferrofluids are made of spherical particles and usually prepared and manipulated by slowly-varying magnetic fields.  The fluctuating part of higher-order magnetic moments are supposed to be of little relevance to macroscopic properties we are concerned with, even if the uniaxial symmetry is not strictly preserved~\cite{Ilg2002magnetoviscosity}. Then it is plausible to approximate the relevant  part of $\mathcal{F}_{int}$ by $\mathcal{F}_0$ that depends only on the magnetization.  Thus,  $\phi_{int}$  can be approximately expressed in terms of the excess effective field as in Eq.~(13).  If this does not work, it would imply the emergence of new microstructures or new slow variables that are appreciably coupled to magnetization dynamics and the theory has to be revised.

Eq.~(17) leads to the following macroscopic relation:
\begin{equation}
M_{eq} = \widetilde{L}(H_s + H^{exc}_{eq}),
\end{equation}  
where $\widetilde{L}(x) = \rho \mu L(\mu x/ k_B T)$ is the scaled Langevin function, with $L(x)= \coth(x)-1/x$.
Now,  denote the equilibrium magnetization equation of state (MEOS) by $M_{eq} = \widetilde{G}(H_s)$,  with $\widetilde{G}$ a function
determined either approximately by analytical models or accurately by experimental measurements.
Then Eq.~(18) can be rewritten as
\begin{equation}
{H}^{exc}(M_{eq}) = \widetilde{L}^{-1}(M_{eq}) - \widetilde{G}^{-1}(M_{eq}).
\end{equation}  
Since $M_{eq}$ can be any allowed number between $0$ and  $\rho \mu$,  the non-equilibrium excess effective field is uniquely determined:
\begin{equation}
\bm{H}^{exc}= \left[\widetilde{L}^{-1}(M(t)) -\widetilde{G}^{-1}(M(t))\right] \widehat{\bm{m}}_t \equiv \bm{H}^L_e(t) -\bm{H}_e(t),
\end{equation}  
with  $\bm{H}_e(t) \equiv \widetilde{G}^{-1}(M(t))\widehat{\bm{m}}_t$ the non-equilibrium thermodynamic effective field conjugate to $\bm{M}(t)$ and
$\bm{H}^L_e(t) \equiv \widetilde{L}^{-1}(M(t))\widehat{\bm{m}}_t$  is the auxiliary Langevin effective field.  Note that for noninteracting ferrofluids  $\bm{H}^{exc}=0$.
Furthermore, with DDIs switched on, the external magnetic field required to prepare a state with a definite magnetization is always smaller than that with DDIs switched off.
Thus, we always have $H^L_e(t) \ge H_e(t)$.

This completes derivation of the DEFM proposed in Ref.~\citenum{FangSM2020}.  It involves a minimum assumption that
the relevant part of $\mathcal{F}_{int}$ is a function of the instantaneous magnetization.  This seems quite reasonable for ferrofluids with typical concentrations and interaction strengths, with magnetization the only relevant slow variable.   Then the excess chemical potential can be expressed in terms of the excess effective field.  The latter can be uniquely determined from equilibrium magnetization curve,  without introducing any adjustable parameters.  Given the quantitative agreement between its theoretical predictions and BD simulations,  I believe the DEFM faithfully describes the dynamics of interacting ferrofluids typically encountered.

In Ref.~\citenum{FangSM2020}, ferrofluids under a magnetic field are considered, with $v_{ext}(\bm{e}, t) = -\mu \bm{e}\cdot \bm{H}(t)$, where the external magnetic field $\bm{H}(t)$ varies sufficiently slowly so as not to hamper the adiabatic approximation.
The projection operator technique~\cite{Zwanzig:1961, Grabert:1982} was employed to coarse-grain Eq.~(15) by treating $\bm{M}(t)$ as the only relevant slow variable.  The magnetization relaxation equation is obtained as
\begin{equation}
\tau_r \frac{d\bm{M}}{dt} =\frac{M}{H^L_e} \delta\bm{H}_{\parallel} + \frac{1}{2}\left(3\chi_L -\frac{M}{H^L_e}\right)\delta\bm{H}_{\perp} + \int_0^t \bm{\Gamma}\left[t, s; \bm{H}(t), \bm{M}(s)\right]\cdot \delta\bm{H}(s),
\end{equation}  
where $\tau_r = 1/2 D^{*}_r$ is the characteristic orientational relaxation time for a dressed particle and  on the right hand side the argument ``$t$" is suppressed for brevity, except for the last term.  The subscripts ``$\parallel$" and ``$\perp$" denote components of the instantaneous thermodynamic driving force $\delta \bm{H}(t)\equiv\bm{H}(t)-\bm{H}_e(t)$ parallel and perpendicular to $\bm{M}(t)$, respectively.  $\bf{\Gamma}$ is a $3 \times 3$ memory matrix generally depending on the magnetization in the past history.  It describes the coupling of collective reorientation  flux to fast degrees of freedom such as the fluctuating part of higher-order moments.   I remark that, to reduce Eq.~(12) to Eq.~(15), we have assumed only $\mathcal{F}_0$,  the uniaxial part of $\mathcal{F}_{int}$, contributes to the interaction-induced microscopic torque experienced by a representative particle.  Nevertheless, the SE (15) still includes contributions of fluctuating higher-order moments  to the torques resulted from thermal motion or external fields. The memory kernel in Eq.~(21) arises from non-uniaxial corrections to the single-particle ODF evolving in time. If particle clustering is insignificant, the external potential varies slowly, and magnetization remains the sole slow variable, then $\bf{\Gamma}$ decays quickly and the memory effect on macroscopic magnetization dynamics can be discarded.   Then we obtain~\cite{FangSM2020}
\begin{equation}
\tau_r \frac{d\bm{M}}{dt} =\frac{M}{H^L_e} (\bm{H}-\bm{H}_e)_{\parallel} + \frac{1}{2}\left(3\chi_L -\frac{M}{H^L_e}\right)(\bm{H}-\bm{H}_e)_{\perp},
\end{equation}  
which is called the generic magnetization relaxation equation (GMRE).  Notably, this GMRE may still be improved by including the memory effect via a  phenomenological kernel, e. g., via a delta-function or exponential function.  It may lead to improved prediction of the anisotropic magnetization relaxation times for a strongly interacting ferrofluid
under a finite bias magnetic field,  where particle clustering along field direction can retard relaxation along the longitudinal but not transverse
direction~\cite{FangSM2020}.

The GMRE well produces previous perturbative models as low-order approximations.  It is readily interpreted in the framework of linear irreversible thermodynamics. Nevertheless,  in general the transport coefficients are nonlinear functions of instantaneous magnetization characterizing the evolving non-equilibrium microstructure.  They have to be determined from non-equilibrium statistical mechanics, as I have done here (see ESI of Ref.~\citenum{FangSM2020}).  In the GMRE (22), it is somehow surprising that the transport coefficients are of  universal, model-independent form and fully determined by $M(t)/\widetilde{L}^{-1}(M(t))$, without involving the MEOS.   The latter, usually material dependent, is only needed to determine $\bm{H}_e$ appearing in the non-equilibrium driving force.  Importantly, even if the MEOS  is not analytically known and does not fit any existing models,  we can still employ the experimentally measured equilibrium magnetization curve to determine $H_e(t)$ as a function of $M(t)$ and use GMRE to predict the magnetization dynamics accurately.

\section{DEFM for Inhomogeneous Ferrofluids and the Demagnetization Effect}
In Sec. II I have considered a homogeneous and unbounded ferrofluid, for which there is no separation between mesoscopic and macroscopic spatial
scales.  However, a real ferrofluid sample is always of finite size and often macroscopically inhomogeneous,  due to applying of a nonuniform magnetic field or flow deformation, or coupling with its surroundings. Is the single-particle SE (15) still valid for describing the mesocopic orientational dynamics (for a volume that is macroscopically small but mesoscopically large)?

If a colloidal dispersion is far away from the critical point and  particle interactions are of short-range nature, the answer is trivially true  because there are no interactions between particles belonging to distinct macroscopic locations.  However, for a ferrofluid in which particles interact via long-range DDIs,
the answer is by no means obvious.   Let me analyze the problem in detail.

I first divide the sample into a number of cells,
denoted by ${\mathcal C}_k$, with $k=0, 1, 2, ..., N_c$.  As usual, the cell size is chosen to be much larger than inter-particle correlation length but much smaller than the length scale characterizing macroscopic inhomogeneity.  Inside each cell mesoscopic homogeneity is assumed, so the nonadiabatic effects on rotational dynamics solely arise from temporal
nonlocality and are captured by renormalizing the rotational self-diffusion coefficient.   A vectorial variable $\bm{R}$ is introduced to parameterize the position space on macroscopic scale.  With each cell macroscopically small, the location of ${\mathcal C}_k$ is specified by  $\bm{R}=\bm{R}_k$.  A macroscopically smooth external magnetic field, denoted by $\bm{H}_{ext}(\bm{R}, t)$,  is applied to the ferrofluid sample,
For a cell ${\mathcal C}_k$, I denote $\rho_k$ the number of particles per unit volume and $W_k(\bm{e}, t)$ the ODF.
Let us focus on a tagged particle, which is oriented along $\bm{e}$ and belongs to ${\mathcal C}_0$.  Without losing generality we may assume $\bm{R}_0=0$.
For each untagged particle in ${\mathcal C}_0$,  a  mesoscopic distance vector $\bm{r}'$ is introduced to describe its position relative to the tagged one, along with
$\bm{e}'$ characterizing its orientation.  The anisotropic part of  potential energy is given by
\begin{equation}
v^{dd}(\bm{r}', \bm{e},\bm{e}') = - \mu^2 \bm{e} \cdot \bm{\mathcal{D}}_2(\bm{r}') \cdot \bm{e}',
\end{equation}  
with $\bm{\mathcal{D}}_2$ the dipole tensorial function defined by
\begin{equation}
\bm{\mathcal{D}}_2(\bm{r}) = (3 \widehat{\bm{r}}\widehat{\bm{r}} - \bm{I}) r^{-3},
\end{equation}  
where $\widehat{\bm{r}}\equiv \bm{r}/r$ and $\bm{I}$ is the identity tensor.   Due to the long-range nature of DDIs, particles belonging to different cells can still interact with each other.   The excess chemical potential of the tagged particle, denoted by $\phi_{exc}$, is contributed by both intra- and inter-cell interactions.   Thus, in general $\phi_{exc}$ is not expected to be a local functional of $W_0$.

To proceed, let me analyze different contributions to $\phi_{exc}$ more carefully.   The contribution due to particles belonging to $\mathcal{C}_k$ ($k\ne 0$) can be written as
\begin{equation}
\phi_{exc}^{(k)} =  \rho_k \Delta V_k  \oint d\bm{e}' W_k(\bm{e}', t) v^{dd}(\bm{R}_k, \bm{e}, \bm{e}')
             =   \Delta V_k \mu \bm{e}\cdot \bm{\mathcal{D}}_2 (\bm{R}_k)\cdot \bm{M}(\bm{R}_k, t)
\end{equation}  
where $\Delta V_k$ is the cell volume for $\mathcal{C}_k$ and I have approximated the inter-particle distance vector by the inter-cell distance vector. The second equality is obtained by integrating over $\bm{e}'$ to yield the k-cell magnetization $\bm{M}(\bm{R}_k, t)=\rho_k \mu \oint d\bm{e}' \bm{e}' W_k(\bm{e}', t)$.
We denote  $\phi_{exc}^{NL} = \sum_{k=1}^{N_c}\phi_{exc}^{(k)}$, with the superscript ``NL" indicating its macroscopically non-local nature.

The contribution due to particles inside $\mathcal{C}_0$ can be separated into two  parts, by decomposing the PCF as $g^{(2)}(\bm{r}, \bm{e}, \bm{e}') = 1+h_2(\bm{r}, \bm{e}, \bm{e}')$, with $h_2$ the total correlation function.  The first part, denoted by $\phi_{exc}^{0I}$,  is simply the independent-particle contribution:
\begin{equation}
\phi_{exc}^{0I} =  \rho_0 \mu^2 \int d\bm{r}'  \oint d\bm{e}' W_0(\bm{e}', t) \bm{e} \cdot \bm{\mathcal{D}}(\bm{r}'),  \cdot \bm{e}',
\end{equation}  
with $\bm{\mathcal{D}}(\bm{r})\equiv \bm{\nabla}\bm{\nabla}(1/r) = \bm{\mathcal{D}}_2(\bm{r})- (1/3)\delta(\bm{r})\bm{I}$ the full dipolar tensor including the singular contribution at $r=0$.  Using $\bm{\mathcal{D}}$ instead of $\bm{\mathcal{D}}_2$ implies particles are treated as independent and the contact contribution should be included. Because the integration of $\bm{\mathcal{D}}_2(\bm{r}')$ over the direction of  $\bm{r}'$  is zero, only the contact term contributes to $\phi_{exc}^{0I}$.
Thus, we simply have
\begin{equation}
\phi_{exc}^{0I}(\bm{e}, t) =  - \frac{1}{3} \mu \bm{e} \cdot \bm{M}(0,t).
\end{equation}  

The second part, denoted by $\phi_{exc}^C$, is an integral over the contribution proportional to $h_2$, arising from neat inter-particle correlations.  Due to
the entanglement of angular and positional variables in the unknown $h_2$,  the integration over $\bm{r}'$ can not be carried out.  I rewrite it in the form
\begin{equation}
\phi_{exc}^{C}(t) = - \mu \bm{e} \cdot \bm{H}^{Cor}(t),
\end{equation}   
where the correlation-induced effective field is defined  by
\begin{equation}
\bm{H}^{Cor}(t) = \rho_0 \mu  \int d\bm{r} \oint d\bm{e}' h_2(\bm{e}, \bm{e}', \bm{r}) \mathcal{D}(\bm{r})\cdot \bm{e}' W_0(\bm{e}',t).
\end{equation}   

Now, adding $\phi_{int}^{NL}$ and $\phi_{exc}^{0I}$ together yields the total independent-particle contribution. Because these cells are macroscopically small, summation over cells can be replaced by integration.  We have
\begin{equation}
\phi_{exc}^{NL}+ \phi_{exc}^{0I} = -\mu \bm{e}\cdot  \bm{H}_{ind}(t),
\end{equation}   
where $\bm{H}_{ind}(t)$ is the independent-particle effective field and given by
\begin{equation}
\bm{H}_{ind}(t) = \int_V d\bm{R}\, \bm{\mathcal{D}}(\bm{R})\cdot \bm{M}(\bm{R},t).
\end{equation}   
Notably, $\bm{H}_{ind}(t)$ is of macroscopic nature and readily identified as the demagnetizing field at $\bm{R}=0$ obeying macroscopic magnetostatics~\cite{Gubbins2011theory}.

It is interesting to recall that the macroscopic field $\bm{H}_{\mu0}(\bm{R})$ due to a magnetic dipole $\mu\bm{e}$ at $\bm{R}=0$ has the form~\cite{Gubbins2011theory}
\begin{equation}
\bm{H}_{\mu0}(\bm{R}) = \mu \bm{e}\cdot \mathcal{D}_2(\bm{R}) - \frac{1}{3}\mu\bm{e} \delta(\bm{R}),
\end{equation}  
with the first term the familiar long-range one and the second the contact contribution.
Hence another expression is obtained for the independent-particle contribution to the excess chemical potential:
\begin{equation}
\phi_{exc}^{NL}+ \phi_{exc}^{0I} = - \int d\bm{R}\, \rho(\bm{R}) \oint d\bm{e}' W_{R}(\bm{e}',t) \bm{e}'\cdot \bm{H}_{\mu0}(\bm{R}).
\end{equation}  
It is easy to check that Eqs.~(33) and ~(30) are identical, as a consequence of Newton's third law.

Now, I denote $\bm{H}_{mw}(t)\equiv  \bm{H}_{ext}(0, t) +\bm{H}_{ind}(t)$ the Maxwell magnetic field at $\bm{R}=0$.  For the rotational dynamics of particles in $\mathcal{C}_0$, we have
\begin{equation}
\frac{\partial W_0(\bm{e}, t)}{\partial t} = \frac{D^{*}_{r0}}{k_B T} \widehat{\mathcal R}_e \cdot W_0 (\bm{e}, t) \widehat{\mathcal R}_e \left[
 k_B T \,  \ln W_0(\bm{e}, t) -\mu \bm{e}\cdot\bm{H}_{mw}(t) - \mu \bm{e}\cdot\bm{H}^{cor}(t) \right],
\end{equation}  
where $D^{*}_{r0}$ is the effective rotational diffusion coefficient for particles inside $\mathcal{C}_0$.  Usually it depends on the local density $\rho_0$ and the DDI strength.  Clearly, the index ``0" can be replaced by $k=1, ..., N_c$ for other macroscopic cells as well.

$\bm{H}^{Cor}$ can be easily determined if inter-particle correlations purely arise from the excluded volume effect.  The internal field inside a particle located at $\bm{r}'$ inside $\mathcal{C}_0$ is opposite to the direction of its magnetic moment, as signalled in  the negative sign of the contact field $\bm{H}^c_{e'}(\bm{r}') = -(1/3)\mu \bm{e}'\delta(\bm{r}')$.  The total microscopic field sensed by the tagged particle at $\bm{r}'=0$ should be the Maxwell field $\bm{H}_{mw}$ excluding the contact field produced by all other particles in $\mathcal{C}_0$.   Therefore,  we have
\begin{equation}
\bm{H}^{Cor} (t) = - \rho_0 \int d\bm{r'} \oint d\bm{e}' W(\bm{e}', t) \bm{H}^c_{e'}(\bm{r}') = \frac{1}{3} \bm{M}(t).
\end{equation}  
Interestingly, this is just the mean-interaction field obtained in the celebrated Weiss model, appropriate for ferrofluids with weak orientational correlations~\cite{Felderhof2003mean, Ilg2005magnetoviscosity}.  In fact, it exactly corresponds to the zeroth-order approximation (with respect to $\lambda$) to the PCF~\cite{Ivanov2001magnetic}, which is just the isotropic PCF for a corresponding hard-sphere fluid.

However, in general $\bm{H}^{Cor}$ can be a complicated function of $\bm{M}(t)\equiv \bm{M}(0, t)$ and other higher-order moments of $W_0$.  To progress, I follow the DDFT scheme again.  Under the adiabatic approximation, $- \mu \bm{e} \cdot \bm{H}^{Cor}$ can be expressed as the functional derivative of correlation-induced free energy with respect to
$W(\bm{e}, t)$.   Now, similar to what is argued in Sec.II, we may assume  $\bm{H}^{Cor}$ is a function of $\bm{M}(t)$ alone.  Following the same procedure we can explicitly determine
\begin{equation}
\bm{H}^{Cor}(t) = \left[\widetilde{L}^{-1}(M(t))- \widetilde{G}^{-1}(M(t))\right] \bm{M}(t)/M(t),
\end{equation}  
where $\widetilde{G}$ is a function describing the dependence of equilibrium magnetization on the Maxwell magnetic field.  For a homogeneous ferrofluid sample subject to a uniform static magnetic field $H_{ext}$, the equilibrium magnetization can be described  either by $M_0= \widetilde{G}_a(H_{ext})$ or by $M_0= \widetilde{G}(H^0_{mw})$, where $H^0_{mw}$ is the sum of $H_{ext}$ and the demagnetizing field.  We may call $M_0= \widetilde{G}_a(H_{ext})$  the apparent MEOS and $M_0= \widetilde{G}(H^0_{mw})$  the canonical or intrinsic MEOS,  both of which can be employed to describe the equilibrium magnetization curve for a specific sample. Nevertheless, the latter is usually adopted because $\widetilde{G}$ other than $\widetilde{G}_a$ is independent of the sample shape and its surroundings, thereby characterizing intrinsic material properties.  For a nonuniform ferrofluid,  the relation between $M_0$ and $H_{ext}$ becomes nonlocal and only the canonical MEOS is appropriate to describe the local thermodynamic state.

In Eq.~(34) $\bm{H}^{Cor}(t)$ is fully determined by $\bm{M}(0, t)$, whereas $\bm{H}_{mw}(t) $ is a local magnetic field at $\bm{R}=0$ available to direct experimental measurements and satisfying macroscopic Maxwell equations.  Thus,  $W_0 (\bm{e}, t)$ seems to be decoupled from the ODF's belonging to other cells and it is tempting to say the mesoscopic dynamics described  by Eq.~(34) is of local nature.  However, due to the demagnetizing contribution, $\bm{H}_{mw}(t)$ implicitly depends on the instantaneous magnetization of all other cells.  Via Eq.~(34), a two-scale parallel algorithm may be designed to iteratively  simulate mesoscopic orientational dynamics throughout the sample,  with the current value of $\bm{H}_{mw}$ in every cell determined from the magnetization of all cells obtained from the preceding time step.

For a macroscopically uniform ferrofluid (far away from boundaries), the rotational dynamics described by Eq.~(15) (supplemented by Eq.~(20)) and Eq.~(34) (supplemented by Eq.~(36)) are equivalent.  In this case the demagnetizing field is uniform and given by $\bm{H}_{ind}= - \alpha \bm{M}$, with $\alpha$ the demagnetizing factor depending on the sample shape and magnetic boundary conditions.  Then we have ${H}_{ind}+ {H}^{Cor}= - \alpha M + \widetilde{L}^{-1}(M)- \widetilde{G}^{-1}(M) = \widetilde{L}^{-1}(M)-\widetilde{G}_a^{-1}(M)= {H}^{exc}$, thereby proving the equivalence of Eqs.~(15) and (34) for this special case.
Nevertheless, by decomposing the total microscopic field sensed by a tagged particle into the sum of
$\bm{H}_{mw}$ and $\bm{H}^{Cor}$, the SE (34) is favorable because it is physically more transparent and applies to real finite-size samples.  For example, for a ferrofluid sufficiently dilute and weakly interacting, naively setting $\bm{H}^{exc}=0$ in Eq.~(15) implies
completely neglecting both short-range (intra-cell) and long-range (inter-cell) interactions.  However, neglecting the latter means discarding demagnetization effect, which would lead to unphysical results for a real bounded sample.  On the other hand,  employing Eq.~(34) with $\bm{H}^{Cor}=0$ still properly accounts for macroscopic demagnetization effect.   Hence we see the important difference between ``interaction-free" and ``correlation-free".   Only the latter represents a qualitatively correct zeroth-order description appropriate for a dilute (weakly correlated) system with long-range interactions.

Interestingly, for a uniform sample of spherical shape or cylindrical shape with hight-to-diameter aspect ratio equal to $1$, we have the demagnetizing factor  $\alpha=1/3$.  So, if  Eq.~(35) (Weiss model) is employed to approximate $\bm{H}^{Cor}$, valid for weakly correlated ferrofluids, then we have $\bm{H}_{ind}+\bm{H}^{Cor} \approx 0$.   Hence, the total microscopic field experienced by a representative particle is just $\bm{H}_{ext}$,  as if it is isolated from other particles and  solely driven by the external magnetic field.
Therefore, we may use the MRSh equation~\cite{MRSh:1974} to describe its magnetization dynamics.   In fact,  by carefully designing the sample shape so as to counterbalance $\bm{H}_{ind}$ and $\bm{H}^{Cor}$,  a group~\cite{Embs2000measuring} has managed to measure the rotational viscosity of a dilute ferrofluid,  which compares well with
the prediction based on the non-interacting MRSh model neglecting both inter-particle correlations and the demagnetizing field.

Finally, for a monodisperse ferrofluid,  whether it is homogeneous or inhomogeneous,  we can rewrite the DEFM in the following form:
\begin{equation}
2 \tau_r \frac{\partial W (\bm{e}, t)}{\partial t} = \frac{1}{k_B T} \widehat{\mathcal R} \cdot W(\bm{e}, t) \widehat{\mathcal R} \left[
 k_B T  ln W(\bm{e}, t) -\mu \bm{e}\cdot \left(\bm{H}_{mw}(t)+ \bm{H}^L_e(t)- \bm{H}_e(t)\right) \right].
\end{equation}   
The corresponding GMRE is of Markovian nature and given by
\begin{equation}
\tau_r \frac{d\bm{M}}{dt} =\frac{M}{H^L_e} (\bm{H}_{mw}-\bm{H}_e)_{\parallel} + \frac{1}{2}\left(3\chi_L -\frac{M}{H^L_e}\right)(\bm{H}_{mw}-\bm{H}_e)_{\perp}.
\end{equation}  
In Eq.~(38) the external magnetic field enters as part of the local Maxwell field, whereas the local demagnetizing field depends on magnetization at all macroscopic locations.  All other relevant quantities, including the transport coefficients,  only depend on the instantaneous local magnetization.

\section{DEFM for polydisperse ferrofluids}
Now consider a colloidal suspension consisting of $N$ particles, which can be divided into $n$ species based on different hydrodynamic diameters or magnetic moments.
Particles belonging to different species become distinguishable, but those belonging to the same species are still indistinguishable.  For simplicity the suspension is assumed homogeneous in the position space, though it can be generalized to the inhomogeneous case by following Sec. III.  In this section,  $\rho$, the particle number per unit volume, is a constant.

The starting point is still the $N$-particle SE (3), without including HIs.
After integrating out all coordinates of $N-1$ particles except one particle belonging to
a species labelled by $k \in \{1,...,n \}$,  a dynamic equation is obtained for the k-species single-particle density.  The latter can be written as
$p_k \rho W^{(k)}(\bm{e}_k, t)$, with $p_k$ the particle number fraction and $W^{(k)}$ the normalized single-particle ODF for k-species.
Similar to Eq.~(7), for each $k$, we have the following single-particle  relaxation equation:
\begin{equation}
\frac{\partial W^{(k)}(\bm{e}_k, t)}{\partial t} = \frac{D^{(k)}_{0r}}{k_B T} \,\widehat{\mathcal R}_k W^{(k)}(\bm{e}_k, t) \cdot \widehat{\mathcal R}_k  \left[k_B T \ln{W^{(k)}(\bm{e}_k, t)} +  v^{(k)}_{ext}(\bm{e}_k, t) + \phi^{(k)}_{int}(\bm{e}_k, t) \right],
\end{equation}  
where $\widehat{\mathcal R}_k \equiv \bm{e}_k \times \frac{\partial}{\partial \bm{e}_k}$ and $D^{(k)}_{0r}$ is the rotational diffusion coefficient for a k-particle.
In Eq.~(39) $v^{(k)}_{ext}$ is the external potential acting on the orientational degrees of freedom of a k-particle.  $\phi^{(k)}_{int}$ is the non-equilibrium excess chemical potential due to interactions of a k-particle with all other particles.  With $v^{(kl)}_2$ the interaction energy between a k-particle and another particle belonging to l-species,
we have
\begin{equation}
\phi^{(k)}_{int}(\bm{e}_k, t)= \sum^n_{l=1} p_l \rho \int d\bm{r} \oint d\bm{e}' g^{(kl)}(\bm{r}, \bm{e}, \bm{e}', t)  W^{(l)}(\bm{e}', t) v^{(kl)}_2(\bm{r}, \bm{e}, \bm{e}'),
\end{equation}   
in which $g^{(kl)}$ is the PCF for two particles belonging to $k$- and $l$-species, respectively.

To close the BBGKY-like hierarchies to the leading order, a time scale coarse-graining can be carried out by following DDFT.  The equilibrium closure relations can be utilized to  decouple Eq.~(39) from PCFs.  At any fixed time $t$ on the coarse-grained time scale, we can simply replace $g^{(kl)}(\bm{r}, \bm{e}, \bm{e}', t)$
by its counterpart $\widetilde{g}^{(kl)}_t(\bm{r}, \bm{e}, \bm{e}')$ for
the equilibrium reference system  in a state specified by the same set of  ODF's $\widetilde{W}^{(k)}_t(\bm{e}_k)= W^{(k)}(\bm{e}_k, t)$ ($k=1,...,n$).  Such an equilibrium state
can always be prepared by a set of external potentials denoted by $\widetilde{v}^{(k)}_t(\bm{e}_k)$ ($k=1,...,n$).

For the reference system in equilibrium, denoting $\mathcal{F} \left[p_k \rho \widetilde{W}^{(k)}_t (\bm{e}_k); k=1,...,n \right]$  as the total Helmholtz free energy functional,  the $k$-species ($k=1, ..., n$) ODF satisfies~\cite{Archer2004ddft}
\begin{equation}
\frac{\delta \mathcal{F} \left[p_k \widetilde{W}^{(k)}_t; k=1,...,n  \right]}{p_k \delta \widetilde{W}^{(k)}_t}  =\widetilde{v}^{(k)}_t + k_B T \ln\left[\lambda^3_{(k)} \widetilde{W}^{(k)}_t\right] +  \frac{1}{p_k} \frac{\delta \mathcal{F}_{int}}{\delta \widetilde{W}^{(k)}_t}= c^{(k)},
\end{equation} 
with $c^{(k)}$ a constant independent of the orientational coordinates.  In Eq.~(41) $\lambda_{(k)}$ is the thermal de Broglie wavelength and $\mathcal{F}_{int}$ is the excess free energy  arising from all inter-particle interactions.  The latter is a functional of all one-body ODFs  $p_l  \widetilde{W}^{(l)}_t$  ($l=1, ..., n$).

Furthermore, according to the YBG relations for orientational degrees of freedom, we also have

\[ k_B T \widehat{\mathcal R}_k \ln W^{(k)}_t(\bm{e}_k)  = -  \widehat{\mathcal R}_k \widetilde{v}^{(k)}_t(\bm{e}_k) \]
\begin{equation}
 - \sum^n_{l=1} p_l \rho \int d\bm{r} \oint d\bm{e}' \widetilde{g}^{(kl)}_t (\bm{r}, \bm{e}_k, \bm{e}')  \widetilde{W}^{(l)}_t(\bm{e}')\widehat{\mathcal R}_k \widetilde{v}^{kl}_2(\bm{r}, \bm{e}_k, \bm{e}'),
\end{equation}  
which is just the generalized torque balance condition for  a representative $k$-particle.

Combining Eqs.~(41) and (42) to eliminate $\widetilde{v}^{(k)}_t$ and noting Eq.~(40), we obtain the non-equilibrium excess chemical potential for a k-particle:
\begin{equation}
p_k  \phi^{(k)}_{int}(\bm{e}_k, t) = \frac{\delta \mathcal{F}_{int}}{\delta W^{(k)}(\bm{e}_k, t)}.
\end{equation}  
Substituting it into Eq.~(39) leads to a closed set of dynamical equations for $W^{(k)}(\bm{e}_k, t)$  ($l=1, ..., n$).  Apparently, whereas the adiabatic approximation eliminates the explicit dependence of $\phi^{(k)}_{int}$ on PCFs,  the ODF's  for different species are still coupled together in $\mathcal{F}_{int}$.

To progress,  we may define the instantaneous $k$-species sub-magnetization via $p_k \bm{M}_k (t) = p_k \rho \mu_k \oint d\bm{e}\, W^{(k)}(\bm{e}, t) \bm{e}$ and the total magnetization is given by $\bm{M}(t) = \sum_{k=1}^n p_k \bm{M}_k (t)$.
 Similar to what is argued in Sec. II, for typical polydisperse ferrofluids made of spherical particles in equilibrium,  if particle clustering is insignificant,  the structural order  is supposed to be sufficiently described by the total magnetization and the fluctuations of higher-order magnetic moments are irrelevant.  Then $\mathcal{F}_0$, referring to the relevant part of $\mathcal{F}_{int}$ that contributes to deterministic torques on particles,  is to a good approximation a function of the total magnetization.
Alternatively, this may be regarded as an envelope approximation, assuming the incoherent or fast-oscillating part of interaction-induced torque is negligible.
Then  we have, up to an irrelevant constant,
\begin{equation}
\phi^{(k)}_{int}(\bm{e}_k, t) =  \frac{\delta \mathcal{F}_0}{p_k \delta W^{(k)}(\bm{e}_k, t)} = - \mu_k \bm{e}_k \cdot \bm{H}^{exc},
\end{equation}  
where $\bm{H}^{exc}$ is the nonequilibrium excess effective field defined via
\begin{equation}
\bm{H}^{exc} \equiv  - \rho \frac{d \mathcal{F}_0(\bm{M}(t))}{d \bm{M}(t)} \frac{\partial \bm{M}(t)}{p_k \partial\bm{M}_k(t)} \equiv -\rho \frac{d \mathcal{F}_0(\bm{M}(t))}{d\bm{M}(t)}.
\end{equation}   
Remarkably, Eqs.~(44) and (45) imply $\phi^{(k)}_{int}$ for different species can be described by the same excess effective field.    This is consistent with the modified mean field theories~\cite{Ivanov2001magnetic, Ivanov2007magnetic, Ivanov2017modified}, known to be accurate for describing equilibrium properties for a wide range of polydisperse interacting ferrofluids.  In those equilibrium theories, the effects of inter-particle correlations are captured by the difference between a single equilibrium effective field and the applied magnetic field.  The former determines the sub-magnetization for all species in chemical equilibrium with each other.

Furthermore, in a way similar to that in Sec. II, we can explicitly determine $\bm{H}^{exc}$ as a function of $\bm{M}(t)$.  Denoting $\widetilde{G}_p$ as the function specifying the equilibrium magnetization curve,  we have
\begin{equation}
\bm{H}^{exc}(t)= \left[\widetilde{L}_p^{-1}(M(t)) -\widetilde{G}_p^{-1}(M(t))\right] \widehat{\bm{m}}_t,
\end{equation}   
with $\widehat{\bm{m}}_t$ the unit vector along the direction of $\bm{M}(t)$ and $\widetilde{L}$ the scaled polydisperse Langevin function defined via
$\widetilde{L}_p(x)= \rho \sum_k p_k \mu_k L(\mu_k x/k_B T)$.  Notably, Eq.~(46) is of the same form as Eq.~(20) for monodisperse ferrofluids: the non-equilibrium excess effective field is just the difference between the thermodynamic effective field $\bm{H}_e(t) \equiv \widetilde{G}_p^{-1}(M(t)) \widehat{\bm{m}}_t$ and the Langevin effective field $\bm{H}^L_e(t) \equiv \widetilde{L}_p^{-1}(M(t)) \widehat{\bm{m}}_t$.

Now, substituting Eqs.~(44) into Eq.~(39)  we obtain the evolution equation for a $k$-particle ($k=1,..., n$):
\begin{equation}
2 \tau^{(k)}_r \frac{\partial W^{(k)}(\bm{e}_k, t)}{\partial t} = \frac{1}{k_B T} \widehat{\mathcal R}_k \cdot W^{(k)}(\bm{e}_k, t) \widehat{\mathcal R}_k \left[
 k_B T  ln W^{(k)}(\bm{e}_k, t) + v^{(k)}_{ext}(\bm{e}_k, t) - \mu \bm{e}_k \cdot\bm{H}^{exc}(t)  \right].
\end{equation}   
where $\tau^{(k)}_r$ is the characteristic rotational relaxation time for a $k$-particle on a slow time scale where all inter-particle correlations decay off and the adiabatic approximation is good.  Usually we have  $\tau^{(k)}_r > 1/2 D^{(k)}_{0r}$ due to nonadiabatic effects arising from short-time inter-particle correlations.

Thus, we obtain a set of $n$ equations,  each describing the evolution of ODF for a distinct species.  This is the polydisperse DEFM.
Importantly, the ODF's for all species are coupled together because $\bm{H}^{exc}$, as a function of the instantaneous total magnetization, is a functional of all ODF's.
Given the initial ODF's for all species,  the set of evolution equations (47) can be numerically solved by an iterative finite-difference method in time domain.

Now, treating $\bm{M}_k$ as the relevant slow variable, we can manipulate Eq.~(47)  with the projection operator technique and  obtain
\[\tau^{(k)}_r\frac{d\bm{M}_k}{dt} = \frac{M_k}{H^L_e} \delta \bm{H}_{\parallel}(t) +\frac{1}{2}\left [3 \chi^L_k -\frac{M_k}{H^L_e}\right]\delta \bm{H}_{\perp}(t)\]
\begin{equation}
 + \int_0^t  ds \hspace{1 mm} {\bm{\Gamma}}^{(k)}\left[t, s; \bm{H}(t), \bm{M}_k(s), \bm{M}(s)\right]\delta \bm{H}(s),
\end{equation}  
where $\delta \bm{H}(t)=\bm{H}(t)-\bm{H}_e(t)$ is the common thermodynamic driving force for all species and $\chi^L_k = \rho \mu^2_k/ 3 k_B T$ is the Langevin susceptibility for $k$-species.  The subscripts ``$\parallel$" and ``$\perp$" denote components parallel or perpendicular to $\bm{M}_k(t)$, respectively.  ${\bm{\Gamma}}^{(k)}$ is the memory matrix describing the coupling of collective reorientation flux of $k$-particles to other degrees of freedom lying in the subspace orthogonal to $\bm{M}_k$.   However, in contrast to the monodisperse case,  we can not discard the memory effects here, because ${\bm{\Gamma}}^{(k)}$ usually contains slowly-decaying components due to the collective orientational degrees of freedom of other species.  Therefore, the sub-magnetization for different species are coupled in a non-Markovian manner and in practice it is impossible to solve the set of $n$ equations in the form of Eq.~(48).

Still, we may further perform a time scale coarse-graining to wash out memory effects and validate the quasi-equilibrium approximation.
On a time scale slow enough, a polydisperse ferrofluid should also obey the principle of non-equilibrium thermodynamics. The rate of change of $\bm{M}(t)$ should be proportional to $\delta \bm{H}(t)$ with transport coefficients depending solely on $\bm{M}(t)$, irrespective of whether the ferrofluid sample is monodisperse or polydisperse.  Therefore, it is expected that the memory effects are  to regulate the relaxation rates of different species so that all $\bm{M}_k (t)$ become synchronized, rendering $\bm{M}(t)$ the sole and adequate slow variable characterizing the instantaneous thermodynamic state.  Presumably, by time scale coarse-graining the memory-effect term in Eq.~(48) can be absorbed into the left-hand side, leading to a regularization of  $\tau^{(k)}_r$.  Then we have
\begin{equation}
\overline{\tau}_R \frac{d\bm{M}_k}{dt} = \frac{M_k}{H^L_e} \delta \bm{H}_{\parallel}(t) +\frac{1}{2}\left [3 \chi^L_k -\frac{M_k}{H^L_e}\right]\delta \bm{H}_{\perp}(t),
\end{equation}  
where $\overline{\tau}_R$ is the regulated relaxation time for the sub-magnetization  of all species and could be a function of $\bm{M}$.

Taking the population-weighted average of Eq.~(49) leads to the polydisperse GMRE
\begin{equation}
\overline{\tau}_R  \frac{d\bm{M}}{dt}=\frac{M}{H^L_e} (\bm{H}-\bm{H}_e)_{\parallel} +\frac{1}{2}\left [3 \overline{\chi}_L -\frac{M}{H^L_e}\right](\bm{H}-\bm{H}_e)_{\perp},
\end{equation}  
where $\overline{\chi}_L =\sum_{k=1}^n p_k \chi^L_k$ is the averaged Langevin susceptibility. 
By comparing the polydisperse GMRE with the monodisperse one, it is clear that $\overline{\tau}_R$ should be state independent.  Therefore,  $\overline{\tau}_R$ can be identified as the averaged single-particle rotational diffusion time on a time scale sufficiently coarse-grained to wash out the memory effects due to incoherent inter-species coupling.  In fact, by comparing the low-frequency part of DMS  from the mesoscopic [Eq.~(47)] and  macroscopic [Eq.~(50)] equations of motion, respectively, we can obtain a definite  relation between $\overline{\tau}_R$ and $\tau^{(k)}_r$ ($k=1, ..., n$):
\begin{equation}
\overline{\tau}_R = \frac{\sum_{k=1}^n p_k \chi^L_k \tau^{(k)}_r}{\sum_{k=1}^n p_k \chi^L_k}.
\end{equation}  
Interestingly,  it is $\tau^{(k)}_r$ other than the corresponding diffusion coefficients that are meaningfully averaged, to describe macroscopic magnetization dynamics of a polydisperse ferrofluid.   Moreover, $\overline{\tau}_R$ depends on the magnetic moment as
well as hydrodynamic diameter of each species.  However, because $\tau^{(k)}_r$ characterizes rotational diffusion of a dressed rather than bare $k$-particle and involves integrated effects of short-time inter-particle correlations,  it is extremely difficult to determine it from first principles.

Eq.~(50) is the first MRE~\cite{FangSM2020} for polydisperse interacting ferrofluids.  It is expected to play a crucial role in studying magnetization dynamics of real ferrofluids, which are usually prepared and manipulated with slowly-varying magnetic fields.  Remarkably, the polydisperse GMRE  is of the same generic form as its monodisperse counterpart.
Such a universal form  indicates its thermodynamical (quasi-equilibrium) nature.   However, the state-dependent transport coefficients can only be obtained from finer-scale dynamics beyond considerations of non-equilibrium thermodynamics.  In this work they have been explicitly determined by coarse-graining the mesoscopic DEFM.

\section{Conclusions}
In this paper the dynamical effective field model (DEFM)~\cite{FangSM2020} for ferrofluids are derived for homogeneous, inhomgeneous, and polydisperse cases.  This is achieved under the general framework of classical dynamical density functional theory (DDFT).   In concentrated and strongly interacting ferrofluids,  the nonadiabatic effect can be important, leading to modifications of the original DDFT.  For ferrofluids that is mesoscopically homogeneous, the major nonadiabtic effect arises from temporal nonlocality or short-time  inter-particle correlations.  This can be approximately captured via a delta-function memory kernel~\cite{NotePower20201109}, leading to renormalization of single-particle diffusivity.  Hence, the original ensemble of bare particles are mapped to an ensemble of dressed particles,  satisfying the effective single-particle equation of motion  in the same form as the original DDFT.

It is further assumed the relevant part (contributing to the total microscopic torque on a representative particle) of excess free energy is a function of the instantaneous magnetization.  Then the chemical potential due to particle interactions can be characterized by an excess effective field, which can be explicitly determined by the equilibrium magnetization curve.   This results in the DEFM, easily implemented but still accurately describing the dynamics of typical ferrofluids.  By further coarse-graining the DEFM to
macroscopic time scale,  a generic magnetization relaxation equation (GMRE) can be obtained,  with state-dependent transport coefficients explicitly given. The GMRE enables us to study magnetization dynamics in regimes far from equilibrium, even if the equilibrium magnetization curve is only empirically known.

For macroscopically inhomogeneous ferrofluid sample, it is shown that the DEFM is more appropriately formulated in terms of local Maxwell field and correlation-induced excess field.
The demagnetizing field naturally emerges from microscopic derivations and is the source implicitly coupling the rotational dynamics at different macroscopic locations.
The dynamics of the whole sample can be obtained by solving the DEFM equations via a two-scale algorithm.

For polydisperse ferrofluids, mesoscopic rotational dynamics of a given species couples with that of other species.  The corresponding DEFM accounts for this, with the excess chemical potential determined as a function of the instantaneous total magnetization instead of sub-magnetization of the considered species.
An elegant and analytic expression for the dynamic magnetic susceptibility (DMS) can be obtained from this polydisperse DEFM.    Its accuracy has been demonstrated for some bidisperse ferrofluid samples~\cite{FangSM2020}.  Furthermore, the resulting polydisperse GMRE takes the same form as its monodisperse counterpart, reflecting its thermodynamical consistency.

In applying the DEFM to ferrofluids that may be concentrated and strongly interacting, it is essential to renormalize the bare diffusion coefficient by incorporating the integrated effect of short-time inter-particle correlations.  Due to the phenomenological approach applied to the memory kernel characterizing temporal nonlocality or non-adiabaticity,  the connection between the effective and bare diffusion coefficients is lost.   This will be restored in Paper II,  along with more detailed studies on the effects of dynamic correlations.

\section*{Acknowledgements}
I acknowledge the support from North China University of Water Resources and Electric Power via Grant No. 201803023.

\section*{Data Availability}
The data that supports the findings of this study are available within the article [and its
supplementary material].

\section*{References}
\bibliography{defm1}
\bibliographystyle{phaip}

\end{document}